# BENCH-MARK SOLUTION FOR A PENNY-SHAPED HYDRAULIC FRACTURE DRIVEN BY A THINNING FLUID

### Alexander M. Linkov

*Institute for Problems of Mechanical Engineering, 61, Bol'shoy pr. V. O., Saint Petersburg, 199178, Russia*
*Saint Petersburg State Polytechnical University, Polytechnicheskaya st., 29, Saint Petersburg 195251, Russia*
*e-mail: linkoval@prz.edu.pl*

**Abstract.** The paper presents a solution for axisymmetric propagation of a penny-shaped crack driven by a thinning fluid. The solution to the accuracy of four significant digits, at least, is obtained on the basis of the modified formulation of hydraulic fracture problem by employing the particle velocity, rather than conventionally used flux. This serves to properly organize iterations in the opening after reducing the problem to the self-similar form. Numerical results obtained show relatively small dependence of self-similar quantities (fracture radius, propagation speed, opening, particle velocity, pressure, flux) on the behavior index of a thinning fluid. The results provide bench marks for the accuracy control of truly 3D simulators and they serve for assigning an apparent viscosity when simulating the action of a thinning fluid by replacing it with an equivalent Newtonian fluid.

## 1. INTRODUCTION

The importance of the solution to the problem for a penny-shaped hydraulic fracture (HF) has been explained in the paper by Savitski & Detournay (2002). These authors gave also a comprehensive review and they obtained accurate solutions for the toughness and viscosity dominated regimes. To the date, their solution is actually the only one which serves for examining the accuracy of various simulators of HF in 3D (e .g. Lecampion et al. 2013). It was also established that in practical applications of HF, the influence of the fracture toughness is negligible.

The results by Savitski & Detournay refer to a Newtonian fluid. Meanwhile, in practice, the fluids used for HF are commonly thinning (e .g. Montgomery 2013). Therefore, it is of value to obtain a solution for the axisymmetric propagation of a penny-shaped fracture driven by a fluid with the power viscosity law. This is the objective of the present paper.

We obtain the solution on the basis of the modified formulation the HF problem suggested and employed by the author with colleagues (see e. g. the reviews in the recent papers by Linkov 2015 and Wrobel & Mishuris 2015). Specifically, using the particle velocity, rather than the flux, opens the possibility to organize fast converging iterations in the opening. The known singularities of the velocity (at the source), of the kernel (at a field point) and of the pressure (at the front) are accounted for by the conventional "subtract-and-add" approach. Then even twenty nodal points are sufficient to obtain solutions to the accuracy of three significant digits. Using 1000 nodal points guaranties four correct digits, at least. We present the self-similar opening at the source and the fracture radius to this accuracy. The distributions of self-similar quantities along a fracture are given in figures. The numerical results for thinning fluids show that, similar to plane-strain problem (Adachi & Detournay 2002), the self-similar quantities do not vary significantly when the behavior index changes from zero (perfectly plastic fluid) to the unity (Newtonian fluid). Assigning an apparent viscosity in the line of the paper (Linkov, 2014) becomes available and the needed equation for it is derived.

## 2. PROBLEM FORMULATION

### 2.1. Formulation in term of physical quantities

The problem is formulated similar to that for a Newtonian fluid (Savitski & Detournay 2002) with the three differences: the viscosity law is of power-type; the speed equation is used instead of the global mass balance, and the equations are written in terms of the particle velocity rather than in terms of the flux. Specifically, for the axisymmetric propagation of a penny-shaped fracture we employ for a fluid:

the continuity equation

$$\frac{\partial w}{\partial t} = -\frac{1}{r}\frac{\partial}{\partial r}(rwv) - q_l,  \tag{2.1}$$

the Poiseuille-type equation

$$v = \left[\frac{w^{n+1}}{\mu'}\left(-\frac{\partial p}{\partial r}\right)\right]^{1/n},  \tag{2.2}$$

and the speed equation

$$v_* = \frac{dr_*}{dt} = \lim_{r \to r_*}\left[-\frac{w^{n+1}}{\mu'}\frac{\partial p}{\partial r}\right]^{1/n}.  \tag{2.3}$$

Herein, $w$ is the fracture opening, $t$ is the time, $r$ is the distance from the source to a considered point, $v$ is the particle velocity, $q_l$ is the term accounting for leak-off into formation, $r_*$ is the distance from the source to the fluid front, $v_*$ is the fracture propagation speed, $\mu' = \theta_n M$, $\theta_n = 2\left(2\frac{2n+1}{n}\right)^n$, $n$ is the behavior index of the fluid, $M$ is its consistency index. For a Newtonian fluid ($n = 1$) with the dynamic viscosity $\mu$, we have $M = \mu$, $\theta_n = \theta_1 = 12$; for a perfectly plastic fluid ($n = 0$) with the shear resistance $\tau_0$, we have $M = \tau_0$, $\theta_n = \theta_0 = 2$.

The elasticity equation and the fracture condition are (Savitski & Detournay 2002):

$$w(\rho r_*) = \frac{8r_*}{\pi E'}\int_0^1 K(\rho,\varsigma)p(\varsigma r_*)d\varsigma,  \tag{2.4}$$

$$K_I = K_{Ic},  \tag{2.5}$$

where $E' = \frac{E}{1-\nu^2}$, $E$ is the Young modulus, $\nu$ is the Poisson's ratio,

$$K(\rho,\varsigma) = \begin{cases} \frac{\varsigma}{\rho}F\left(asin\sqrt{\frac{1-\rho^2}{1-\varsigma^2}},\frac{\varsigma^2}{\rho^2}\right), & \varsigma < \rho \\ F\left(asin\sqrt{\frac{1-\varsigma^2}{1-\rho^2}},\frac{\rho^2}{\varsigma^2}\right), & \varsigma > \rho \end{cases},  \tag{2.6}$$

$F(\varphi, k^2)$ is the elliptic integral of the first kind, $K_I$ is the stress intensity factor (SIF) of the normal traction, which in the axisymmetric case is given by equation

$$K_I = \frac{2\sqrt{r_*}}{\sqrt{\pi}}\int_0^1 \frac{p(\varsigma r_*)}{\sqrt{1-\varsigma^2}}\varsigma d\varsigma.  \tag{2.7}$$

When writing the elasticity equation, the small lag between the fluid front and the fracture contour is neglected. The equation is solved under the boundary condition (BC) of zero opening at the crack tip:

$$w(r_*t) = 0.  \tag{2.8}$$

Note that the BC (2.8) is automatically satisfied by the form (2.4) of the elasticity equation because the kernel (2.6) is zero when $\rho = 1$ ($r = r_*$). We need also to meet the BC of the prescribed influx $q_0(t)$ at the source:

$$2\pi w(0,t)\lim_{r \to 0}(vr) = q_0(t).  \tag{2.9}$$

The BC (2.9) implies that to have finite non-zero pumping rate $q_0(t)$, the particle velocity should be singular as $O(1/r)$ at the source. Then for a finite non-zero opening at the source, the Poiseuille-type equation (2.2) implies that the derivative of the pressure behaves as $O(1/r^n)$. Hence, in the case, of a Newtonian fluid ($n = 1$), the pressure has log-type singularity at the source, while in the case of a thinning fluid ($0 < n < 1$) it behaves as $O(1/r^{1-n})$. Besides, near the front, there is also singularity of the pressure discussed below. These singularities are to be accounted for when developing a numerical method.

The initial conditions (IC) when solving the problem (2.1)-(2.9) are of zero fracture opening and radius at the initial time $t = 0$:

$$w(r, 0) = 0, \qquad r_*(0) = 0. \tag{2.10}$$

The problem consists of solving the system (2.1), (2.2), (2.4) under the IC (2.10) and the BC (2.8), (2.9). The solution has to satisfy the fracture condition (2.5), defining the very possibility of the fracture propagation, and the speed equation (2.3), defining the position $r_*$ of the front as a function of time. The kernel of the elasticity equation is given by (2.6).

## 2.2. Self-similar formulation

It is convenient to exclude the elasticity constants ($E$, $v$) and the consistency index ($M$) from the elasticity and Poiseuille-type equations by a proper re-scaling variables. The latter can be done in various ways, of which, for certainty and simplicity, we use the following. We introduce the normalizing opening $w_n$ and rescaled quantities of opening $w'$, pressure $p'$, leak-off $q'_l$ and pumping rate $q'_0$ defined as

$$w_n = \left(\frac{\mu'}{E'}\right)^{\frac{1}{n+2}}, \qquad w' = \frac{w}{w_n}, \quad p' = \frac{p}{w_n E'}, \quad q'_l = \frac{q_l}{w_n}, \qquad q'_0 = \frac{q_0}{w_n}. \tag{2.11}$$

Note that by the first of (2.11), the normalizing opening $w_n$ has the unusual dimension of time in degree $\frac{n}{n+2}$, and it depends on the both the consistency and behavior indices of a fluid.

With the definitions (2.11), it is sufficient merely to change the quantities in equations (2.1)-(2.10) to the primed ones and to change $E'$ and $\mu'$ to 1. After that, we can represent the solution in separated variables, which are the relative distance $\varsigma = r/r_*$ and the time $t$:

$$r_* = \xi_* t^{\gamma_r}, \quad v_* = \xi_* \gamma_r t^{\gamma_r - 1}, \quad w' = \xi_* W(\varsigma) t^{\gamma_w}, \quad v = \xi_* V(\varsigma) t^{\gamma_r - 1}, \quad p' = P(\varsigma) t^{\gamma_p}, \tag{2.12}$$

The exponents $\gamma_r$, $\gamma_w$ and $\gamma_p$ in (2.12) are chosen to obtain equations entirely in $\varsigma$. To this end, we take the influx and the leak-off term as power functions of time:

$$q'_0 = Q_0 t^{\gamma_q}, \quad q'_l = Q_l(\varsigma) t^{\gamma_w}, \tag{2.13}$$

where $\gamma_q$ is an arbitrary real number ($\gamma_q = 0$ for a constant pumping rate). The choice of the leak-off term in the form of the second of (2.13) serves us to cancel time dependent factors in the continuity equation after changing the spatial coordinate $r$ to the relative distance $\varsigma$ and substitution $w'$ and $v$ from (2.12). The time depending factors cancel for arbitrary $\gamma_w$ and $\gamma_r$. The Poiseuille-type and elasticity equations yield $\gamma_p = -n/(n+2)$ and $\gamma_w = \gamma_r - n/(n+2)$. Then the BC at the source provides the value of $\gamma_r$. Finally the constants are:

$$\gamma_r = V_* = \frac{1}{3}\left(1 + \frac{n}{n+2} + \gamma_q\right), \quad \gamma_w = \frac{1}{3}\left(1 - \frac{2n}{n+2} + \gamma_q\right), \quad \gamma_p = -\frac{n}{n+2}, \tag{2.14}$$

where we also introduced the notation of the self-similar propagation speed $V_*$ used below (actually it equals to $\gamma_r$).

With the variables (2.12), power terms (2.13) and exponents (2.14), the system of equations obtains the almost self-similar form:

$$\frac{d[(\varsigma V_* - V)\varsigma W]}{d\varsigma} = (1 + \gamma_q)\varsigma W - Q_l \tag{2.15}$$

$$V = \left(-W^{n+1}\frac{\partial P}{\partial \varsigma}\right)^{1/n} \tag{2.16}$$

$$V_* = \gamma_r = \lim_{\varsigma \to 1}\left(-W^{n+1}\frac{\partial P}{\partial \varsigma}\right)^{1/n} \tag{2.17}$$

$$W(\rho) = \frac{8}{\pi}\int_0^1 K(\rho,\varsigma)P(\varsigma)d\varsigma \tag{2.18}$$

$$\int_0^1 \frac{P(\varsigma)}{\sqrt{1-\varsigma^2}}\varsigma d\varsigma = K_{Ic}{}'(t) \tag{2.19}$$

$$W(1) = 0 \tag{2.20}$$

$$2\pi\xi_*{}^3 W(0)\lim_{\varsigma \to 0}(V\varsigma) = Q_0, \tag{2.21}$$

where the only equation, containing the time-depending factor $K_{Ic}{}'(t)$, is the fracture condition (2.19), in which

$$K_{Ic}'(t) = \frac{1}{2}\sqrt{\frac{\pi}{\xi_*}}K_{Ic}\,t^{-(\gamma_p+\gamma_r/2)t}. \tag{2.22}$$

Obviously $K_{Ic}'$ also becomes time-independent when either we neglect the fracture toughness ($K_{Ic} = 0$), or when $\gamma_p + \frac{\gamma_r}{2} = 0$. According to (2.14), the latter occurs when $n = n_k = 2(1+\gamma_q)/(4-\gamma_q)$. In particular, for a constant pumping rate ($\gamma_q = 0$), the problem (2.15)-(2.21) is entirely self-similar when $n = 0.5$. Remarkably, the exponent $-(\gamma_p + \gamma_r/2)$ on the r. h. s. of (2.22) may be positive if a thinning fluid has the behavior index in the range $n_k < n < 1$. In this range, the influence of the fracture toughness exponentially grows in time. Below we shall focus on the practically important case of the viscosity dominated regime, for which one can set $K_{Ic} = 0$.

The IC (10) are met at $t = 0$ because both $\gamma_r$ and $\gamma_w$ are positive. We are looking for a solution to the system (2.15)-(2.21) for a fixed, in particular zero, value of $K_{Ic}{}'$.

## 3. METHOD OF SOLUTION

There are specific features of the HF problem when neglecting the lag and fixing the position of the front, explained in details in the paper (Linkov 2015). Specifically, the problem is ill-posed in the Hadamard (1902) sense if trying to solve it as a boundary value problem under a fixed position of the front. In the self-similar formulation (2.15)-(2.21), the front position is fixed by prescribing its self-similar value $\xi_*$. Thus the BC at the source (2.21) cannot be used with $\xi_*$ prescribed. Rather we firstly solve the remaining equations (2.15)-(2.20) and find their solution $W(\varsigma)$, $V(\varsigma)$, $P(\varsigma)$. Afterwards, when having $W(\varsigma)$ and $V(\varsigma)$ known, we insert the value of the product $W(0)\lim_{\varsigma \to 0}(V\varsigma)$ into (2.21) to find the self-similar radius $\xi_*$, which corresponds to the prescribed self-similar influx $Q_0$:

$$\xi_* = \left(\frac{2\pi}{Q_0}W(0)\lim_{\varsigma \to 0}(V\varsigma)\right)^{-1/3}. \tag{3.1}$$

Actually the system (2.15)-(2.20) is solved as a Cauchy problem with two conditions prescribed at the same point, which is the front $\varsigma = 1$. To see it and to obtain asymptotics of the opening and pressure recall that when neglecting the lag, the speed equation through the elasticity equation uniquely defines the opening near the fracture front as a function of merely the propagation speed (Linkov 2015). In particular, for the considered self-similar problem in the case $K_{Ic}' = 0$, the universal asymptotic umbrella follows from the well-known properties of plane-strain elasticity operator (see, e. g. Muskhelishvili, 1953):

$$W(\varsigma) = A_W(1-\varsigma)^\alpha, \quad P(\xi) = -A_W B(\alpha)(1-\varsigma)^{\alpha-1}, \tag{3.2}$$

where $B(\alpha) = \frac{\alpha}{4}\cot[\pi(1-\alpha)]$. The exponent $\alpha$ and the coefficient $A_W$ are found by the substitution (3.2) into the SE (2.17). They are:

$$\alpha = \frac{2}{n+2}, \ A_W = \left[\frac{V_*{}^n}{(1-\alpha)B(\alpha)}\right]^{\frac{1}{n+2}}. \tag{3.3}$$

Furthermore, using (3.2) and the SE (2.17) in the continuity equation (2.15) provides the asymptotics for the particle velocity near the front:

$$V(\varsigma) = V_*[1 - a_V(1 - \varsigma)], \tag{3.4}$$

where $a_V = \frac{1+\gamma_q}{V_*(1+\alpha)} - 1$. When obtaining (3.4), it is assumed that the leak-off term goes to zero near the front as $o((1-\varsigma)^{\alpha+\varepsilon})$ with $\varepsilon > 0$.

From (3.2)-(3.4) we see that the initial (Cauchy) conditions (2.17), (2.20) for the ordinary differential equations (ODE) (2.15), (2.16) completely define the opening and the particle velocity near the point $\varsigma = 1$. Therefore, as mentioned, there is no need in the BC (2.21) at the source when integrating the ODE.

Integration of (2.15) with the conditions (2.17), (2.20) at $\varsigma = 1$, provides the particle velocity explicitly expressed via the opening:

$$V = \varsigma V_* + \frac{1}{\varsigma}\varphi_V(\varsigma) \tag{3.5}$$

with

$$\varphi_V(\varsigma) = \frac{1+\gamma_q}{W(\varsigma)}\int_\varsigma^1 W(\varsigma)\varsigma d\varsigma. \tag{3.6}$$

Thus, when having an opening $W(\varsigma)$, we also have the particle velocity defined by (3.5), (3.6). According to (3.5), the particle velocity, as required by (2.9), is singular at the source as $O(1/r^n)$. Specifically, its asymptotic behavior implies that

$$\lim_{\varsigma \to 0}(\varsigma V) = \varphi_V(0) = \frac{1+\gamma_q}{W(0)}\int_0^1 W(\varsigma)\varsigma d\varsigma. \tag{3.7}$$

This expression for the factor in the asymptotic formula $V(\varsigma) = \varphi_V(0)/\varsigma$ should be accounted for when evaluating the pressure.

By using (3.5) in the Poiseuille-type equation (2.16), we find the derivative of the net-pressure for a given $W(\varsigma)$:

$$-\frac{d(P-P_0)}{d\varsigma} = \frac{V^n}{W^{n+1}}, \tag{3.8}$$

Herein $P_0$ is the value of the pressure at an arbitrary start point $\varsigma_0$ when integrating (3.8). Then for the self-similar net-pressure we obtain

$$P(\varsigma) = P_0 - \varphi_P(\varsigma), \tag{3.9}$$

with

$$\varphi_P(\varsigma) = \int_{\varsigma_0}^\varsigma \frac{V^n}{W^{n+1}}d\varsigma. \tag{3.10}$$

Substitution of (3.9) into the fracture condition (2.19) defines the constant $P_0$:

$$P_0 = K'_{Ic} + \int_0^1 \frac{\varphi_P(\varsigma)}{\sqrt{1-\varsigma^2}}\varsigma d\varsigma. \tag{3.11}$$

With known $P(\varsigma)$, the elasticity equation (2.18) gives a new (iterated) opening $W(\varsigma)$.

The sequence discussed suggests solving the problem by iterations in the opening and particle velocity, which follow the chain (3.5)-(3.11), (2.18) starting from an initial guess $W(\varsigma) = W_{start}(\varsigma)$. For a perfectly plastic fluid ($n = 0$), there is no need to involve equations (3.5), (3.6) on iteration steps, because in this case we have $V^n = 1.0$ on the r. h. s. of (3.8). Then the equations (3.5), (3.6) are employed *after* completing the iterations to find the distribution of the particle velocity.

*Comment.* This natural scheme of iterations is *ad hoc*, because it is essentially based on the form of the elasticity equation (2.4) solved in the opening. Such a favorable analytical form is available merely in the simplest cases of the plane strain problem for a straight or a circular-arc crack and of the considered axisymmetric problem for a penny-shaped crack. For the plane-strain HF problem, such a scheme has been suggested and successfully employed by Wrobel & Mishuris (2015). In general, however, we have the elasticity equation solved in the net-pressure with the operator on its r. h. s. being hypersingular. This strongly complicates using the iterations in the opening if not inverting the hypersingular equation numerically.

After completing the iterations, the self-similar radius $\xi_*$ is found from the equation

$$\xi_* = \left( \tfrac{2\pi}{Q_0} (1 + \gamma_q) \int_0^1 W(\varsigma) \varsigma \, d\varsigma \right)^{-1/3}, \tag{3.12}$$

obtained by substitution of (3.7) into the BC (3.1). Similar to plane-strain HF problems, the form (3.12) of the BC (3.1) at the source may be interpreted as the equation of global mass balance.

Emphasize that for a given power-law influx, in particular for a constant pumping rate ($\gamma_q = 0$), the constant

$$\xi_{*n} = \left( 2\pi (1 + \gamma_q) \int_0^1 W(\varsigma) \varsigma \, d\varsigma \right)^{-1/3}, \tag{3.13}$$

corresponding to the unit influx ($Q_0 = 1$), depends merely on the behavior index $n$ of a fluid. For an arbitrary influx, from (3.13), (3.12) we have:

$$\xi_* = \xi_{*n} \sqrt[3]{Q_0}. \tag{3.14}$$

In terms of the physical values of the fracture radius $r_*$ and the influx $q_0$, the definitions (2.11)-(2.13) and equation (3.14) yield

$$r_* = \xi_{*n} \sqrt[3]{q_0/w_n} \, t^{\gamma_r - \gamma_q/3}, \tag{3.15}$$

where $w_n$ and $\gamma_r$ are defined by the first of (2.11) and (2.14), respectively.

## 4. NUMERICAL RESULTS AND DISCUSSION

The iterative scheme described is used in frames of the finite differences approach. We represent the iterated quantities (self-similar opening, particle velocity and pressure) by their values at nodes of a uniform spatial mesh dividing the interval [0,1] to $N$ equal segments. Integrations over the first half-segment are performed with accounting for the asymptotic equation (3.7). Integrations over the last half-segment are performed with taking into account the asymptotic behavior (3.2)-(3.4) of each of the iterated quantities. This is done by the common "subtract-and-add" method with analytical evaluation of the added integrals. Similar approach is used to account for the log-type singularity of the kernel in the elasticity equation. For the remaining smooth integrands, we employ the trapezoid rule when performing successive integrations in (3.6), (3.10), (3.11) and (2.18).

For certainty, the calculations are performed for an impermeable rock ($Q_l = 0$) and constant pumping rate ($\gamma_q = 0$). The starting opening $W_{start}(\varsigma)$ is prescribed quite roughly by setting it proportional to that of a penny shaped-crack under constant unit pressure: $W_{start}(\varsigma) = k(8/\pi)\sqrt{1 - \varsigma^2}$ with $k$, which we changed to see its influence on the convergence of iterations. It appeared that even large (from 0.01 to 1.7) changes of $k$ did not influence significantly the convergence and the number of iterations needed to have results reproduced on successive iterations. To avoid oscillations, at the end of an iteration we take the nodal openings as average of values obtained on the current and previous iteration. Actually twenty iterations are enough to have reliably reproduced five significant digits of the opening even when the starting opening drastically differs from the solution, in particular, when $k = 0.01$. For $k = 1$, it is sufficient to make ten iterations. For the mesh with $N = 1000$ segments, the time expense for twenty iterations does not exceed a minute on a conventional laptop. Special tests revealed that employing this mesh and number of iterations provided five correct significant digits of the opening. For $N = 20$, the accuracy, although two orders less, is still acceptable, while the time expense is a fraction of a second. Therefore the scheme is quite stable, accurate and time-efficient when accounting for the singularities as explained.

The results, obtained to the accuracy of four correct significant digits, at least, for fluids with various behavior indices, are summarized in Table. It presents the self-similar opening $W(0)$ at the source (the second row) and the constant $\xi_{*n}$, defined by (3.13) (the third row). Note that for a Newtonian fluid ($n = 1$), the values presented in the table agree with those obtained for this case by Savitski & Detournay (2002). Specifically, for $W(0)$ we have 1.7092 against 1.713 of these authors; for $\xi_{*N}$ we have 0.6978 against 0.6976. Figures 1-4 present distributions of the self-similar opening $W$, velocity $V$, pressure $P$ and flux $Q$, respectively. To see distinctly the influence of the behavior index on the particle velocity, Fig. 5 presents the $V(\varsigma)$ distribution beyond the near-source zone.

Table

| $n$ | 0.0 | 0.1 | 0.2 | 0.4 | 0.6 | 0.8 | 1.0 |
|---|---|---|---|---|---|---|---|
| $W(0)$ | 1.6889 | 1.6724 | 1.6617 | 1.6537 | 1.6599 | 1.6784 | 1.7092 |
| $\xi_{*n}$ | 0.7330 | 0.7318 | 0.7296 | 0.7236 | 0.7162 | 0.7076 | 0.6978 |
| $(\xi_{*N}/\xi_{*n})^9$ | 0.6422 | 0.6517 | 0.6696 | 0.7212 | 0.7912 | 0.8820 | 1.0 |

The results show that, similar to the self-similar plane-strain problems (Adachi & Detournay 2002, Linkov 2014), the influence of the behavior index is not great when considering the self-similar quantities. This does not mean that the conclusion is true for physical quantities. Quite contrary, for the latter, the influence of the behavior index is significant and the quantities depend also on the consistency index $M$. Meanwhile, analogously to the plane-strain cases (Linkov 2014), the found dependence of $\xi_{*n}$ on $n$ may serve to compare fluids with various behavior and consistency indices in their hydraulic fracturing action. It also provides an opportunity to assign an apparent viscosity when simulating the action of a thinning fluid by replacing it with an equivalent Newtonian fluid.

Following the line of the paper (Linkov 2014), we assume fluids equivalent in their HF action when under the same pumping rate $q_0$ they produce fractures of the same size $r_*$ for a typical reference time $t_r$ of a treatment. The fracture radius $r_*$ is found by using $\xi_{*n}$ in (3.15). Then equating the radii for two fluids, marked by the subscripts 1 and 2, after some algebra, we obtain that the fluids are equivalent in their HF action when

$$\mu_1' = \left(\frac{\xi_{*n1}}{\xi_{*n2}}\right)^{3(n_1+2)} E' t_r^{n_1} \left(\frac{\mu_2'}{E' t_r^{n_2/n_1}}\right)^{\frac{n_1+2}{n_2+2}}. \tag{3.16}$$

Recall that $\mu' = \theta_n M$ and take the first fluid as a Newtonian fluid ($n_1 = 1$, $\theta_{n1} = 12$) with the dynamic viscosity $\mu_a$. Then (3.16) defines a Newtonian fluid equivalent in its fracturing action to a given fluid with the behavior index $n_2 = n$ and the consistency index $M_2 = M$:

$$\mu_a = \frac{1}{12}\left(\frac{\xi_{*N}}{\xi_{*n}}\right)^9 E' t_r \left(\frac{M\theta_n}{E' t_r^n}\right)^{\frac{3}{n+2}}. \tag{3.17}$$

The values of $(\xi_{*N}/\xi_{*n})^9$ are given in the last row of Table. Notably, in contrast with the analogous result (Linkov 2014) for the Nordgren's (1972) problem, the apparent viscosity $\mu_a$ defined by (3.17) includes neither the regime exponent ($\gamma_q$), nor the magnitude ($Q_0 w_n$) of the pumping rate. Note also that in the limiting case of a perfectly plastic fluid ($n = 0$), for which the influences of the reference time $t_r$ and the elasticity modulus $\sqrt{E'}$ are maximal, the apparent viscosity is proportional to $t_r$ and inversely proportional to $\sqrt{E'}$. In the intermediate case between perfectly plastic and Newtonian fluids, when $n = 0.5$, the dependences are much weaker: the apparent viscosity is proportional to $t_r^{2/5}$ and inversely proportional to $E'^{1/5}$.

We may compare the apparent viscosity defined by (3.17) with the empirical value given as an example in the key-note lecture by Montgomery (2013). The author considered a thinning fluid with the behavior index $n = 0.6$ and the consistency index $M = 0.39$ Pa·s$^{0.6}$. For this fluid, when taking the same typical values of the elasticity modus $E = 2.5 \cdot 10^4$ MPa, Poisson's ratio $\nu = 0.15$ and treatment time $t_r = 10^4$ s as those in the paper (Linkov 2014), equation (3.17) yields $\mu_a = 84$ cps against the value $\mu_a = 81$ cps suggested by Montgomery. The agreement, although a bit accidental, looks perfect.

***Acknowledgement.*** The author gratefully acknowledges the support of the Russian Scientific Fund (Grant # 15-11-00017).

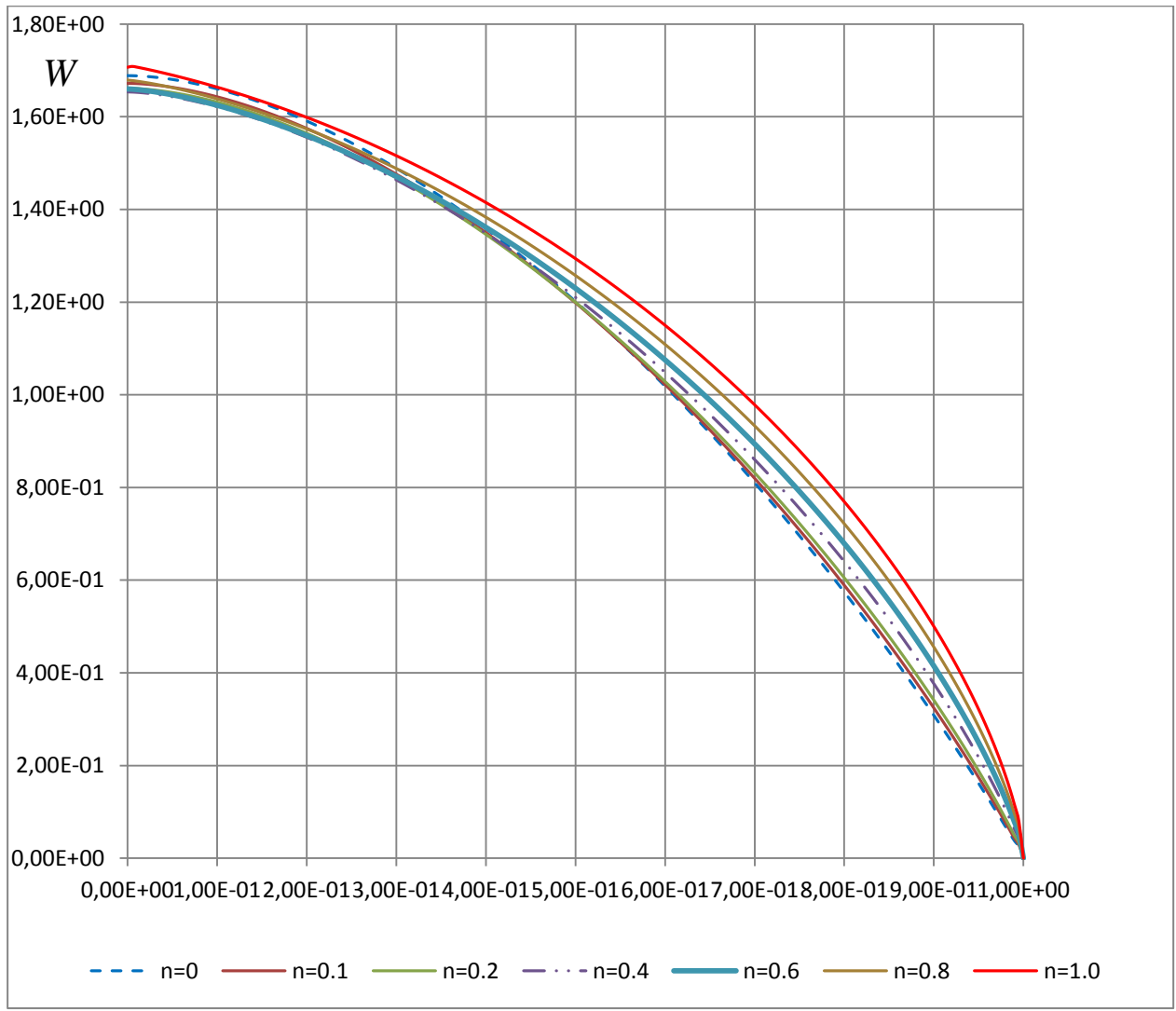

Fig. 1. Distributions of self-similar opening for various behavior indices

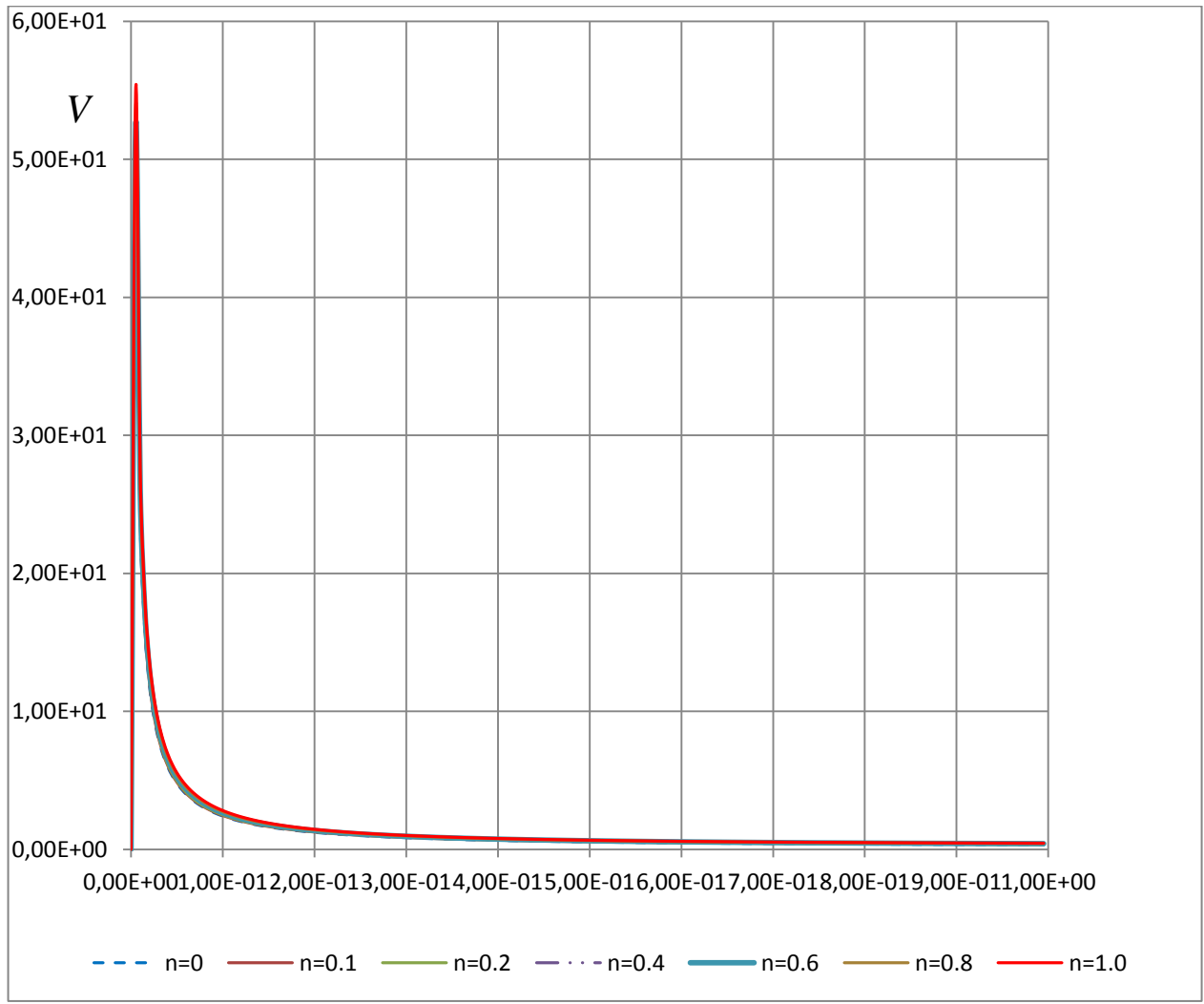

Fig. 2. Distributions of self-similar particle velocity for various behavior indices

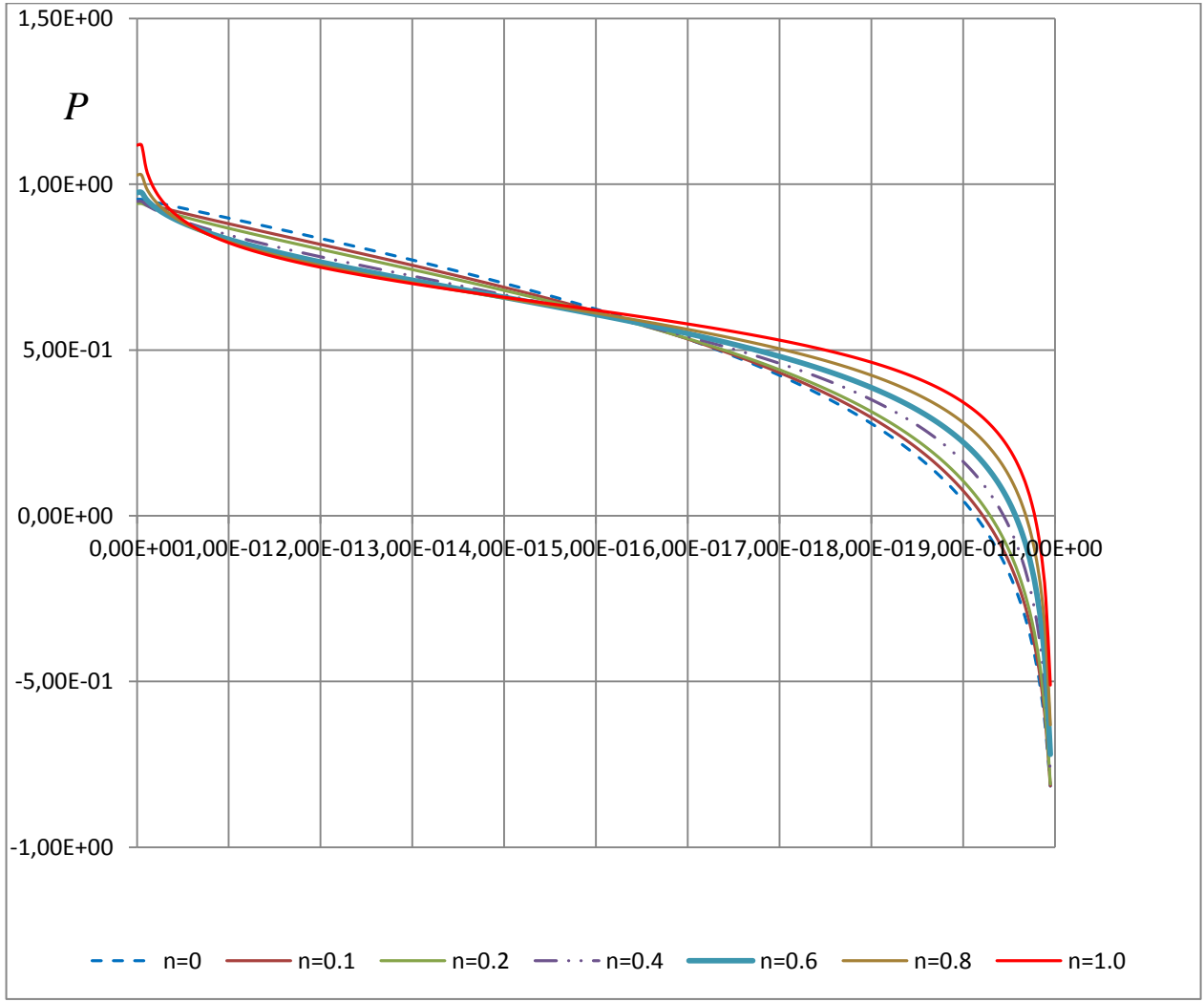

Fig. 3. Distributions of self-similar net-pressure for various behavior indices

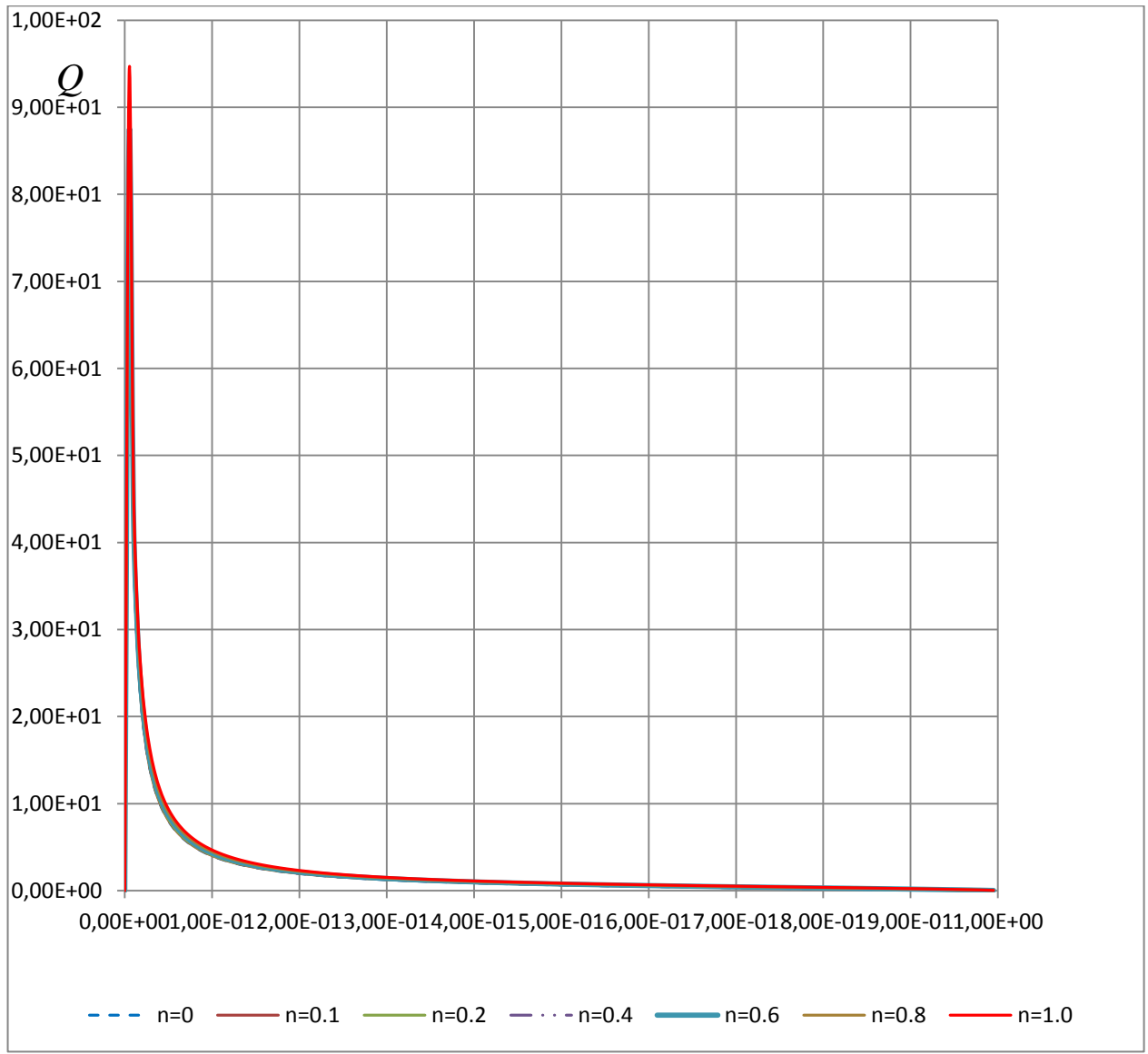

Fig. 4. Distributions of self-similar flux for various behavior indices

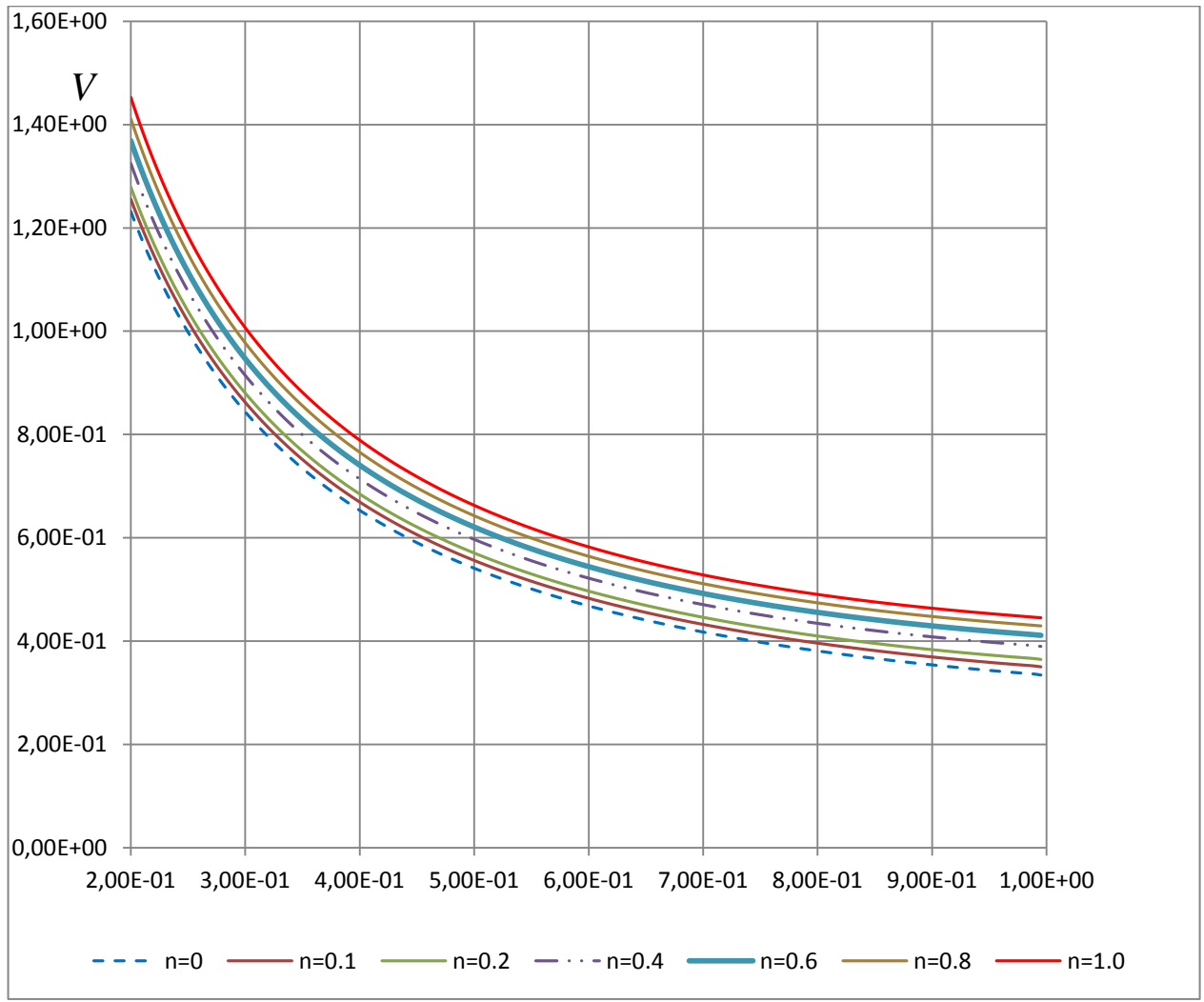

Fig. 5. Distributions of self-similar velocity beyond near-source zone